\documentclass[aps,nofootinbib,superscriptaddress,floatfix,a4paper,showkeys]{revtex4-2}

\usepackage{orcidlink,amssymb,amsmath,hyperref,siunitx,makecell}
\usepackage{enumitem}
\usepackage{ytableau,graphicx,youngtab}
\usepackage{soul}
\usepackage{bm}
\usepackage{multirow}
\usepackage{ulem}
\allowdisplaybreaks
\usepackage{graphicx}
\usepackage{booktabs}
\usepackage{xcolor}
\usepackage{comment}
\usepackage{tikz}
\usepackage{booktabs}

\hypersetup{pdfnewwindow=true,
	colorlinks=true,linkcolor=blue,
	citecolor=blue,filecolor=blue,urlcolor=blue
}

\definecolor{maroon}{RGB}{128,0,0}

\def\IMSc{The Institute of Mathematical Sciences, CIT Campus, Chennai, 600113, India}
\def\HBNI{Homi Bhabha National Institute, Training School Complex, Anushaktinagar, Mumbai 400094, India}
\newcommand{\imsc}{\affiliation{\IMSc}}
\newcommand{\hbni}{\affiliation{\HBNI}}

\newcommand{\beit}{\begin{itemize}}
\newcommand{\eeit}{\end{itemize}}

\newcommand{\bec}{\begin{center}}
\newcommand{\eec}{\end{center}}

\newcommand{\beq}{\begin{equation}}
\newcommand{\eeq}[1]{\label{eq:#1}\end{equation}}

\newcommand{\beqa}{\begin{eqnarray}}
\newcommand{\eeqa}{\end{eqnarray}}

\newcommand{\bet}{\begin{table*}}
\newcommand{\eet}[1]{\label{tb:#1}\end{table*}}

\newcommand{\bef}{\begin{figure}}

\newcommand{\eef}{\end{figure}}

\newcommand{\cref}[1]{{\normalsize\color{refdark}\itshape #1}}

\begin{document}
	
	\title{Sign of Wilson-Fermion Determinant using Contour-Integral Spectral Projection}
	
	\author{Bhabani Sankar Tripathy\orcidlink{0000-0001-7759-2778}}
	\email{bhabanist@imsc.res.in}
	\imsc
	\hbni
	
	\author{M. Padmanath\orcidlink{0000-0001-6877-7578}}
	\imsc
	\hbni
	
	\begin{abstract}
		Key properties of a large sparse operator can often be determined by only a small, spectrally localized subset of its eigenmodes. In lattice QCD with Wilson fermions, the sign of the fermion determinant is determined by the parity of the number of eigenvalues lying on the negative real axis of the non-Hermitian Wilson-Dirac operator. Determination of this signature can be challenging, particularly in parameter regimes where near-zero modes can lead to exceptional configurations. We present a contour-integral-based spectral projection method to count the negative-real eigenvalues of the non-Hermitian Wilson-Dirac operator $D$ that determine the sign of its determinant. The spectral projection is supplemented by an adapted deflation-accelerated solver for shifted linear systems and singular-value exclusions to reliably set spectral bounds and contour boundaries. We demonstrate the robustness of the method through systematic convergence tests, including variations of the contour size, and show its ability to resolve eigenvalues close to the negative real axis. Beyond fermion determinant sign evaluation, the approach provides a systematic means of isolating physically relevant eigenvalues embedded in dense complex spectra, with potential applications to non-Hermitian quantum systems, stability analyses, and large-scale eigenvalue problems.
	\end{abstract}
	\maketitle
	
	
	\section{Introduction}
	Physically important properties of large sparse matrices are often controlled by their low lying eigenmodes, \textit{c.f.} Ref. \cite{Neff:2001zr}. In lattice QCD ensembles assuming Wilson fermion action, such a connection appears in the signature of the fermion determinant $\det(D)$, where $D$ denotes the Wilson-Dirac operator, which encodes the integrated dynamics of Grassmann-valued fermion field. While the $\gamma_5$-Hermiticity ensures that the determinant is real, the positivity of the fermion determinant is not guaranteed for single flavor or mass-nondegenerate Wilson fermions. The $\gamma_5$-Hermiticity property also implies that the sign of Wilson-Dirac fermion determinant is exclusively determined by $N_{-}$, the number of eigenvalues of the Wilson-Dirac operator lying on the negative real axis  \cite{Mohler:2020txx}. Determining this signature can be challenging, particularly in the presence of near zero modes, where the fermion operator is poorly conditioned.
	
	In lattice QCD with a single Wilson fermion flavor \cite{Francis:2018xjd,DellaMorte:2023ylq}, or with more than one dynamical mass-nondegenerate Wilson fermion flavor \cite{RCstar:2022yjz}, a positive definite ({\it pseudo})fermion action for each flavor can be constructed from the square root of $D^\dagger D$, whose fractional power can be efficiently approximated using polynomials \cite{Frezzotti:1998eu} or using rational functions \cite{Clark:2006fx}, with corrections to the approximation accounted for by an exact Metropolis accept-reject step following the approximation \cite{JLQCD:2001ucs}. In lattice QCD ensembles such as those generated by the CLS consortium, the residual rational approximation errors were not accounted for at the simulation stage. In such setups, the accumulated error in the simulated fermionic measure from the rational/polynomial approximations should be corrected by an appropriate reweighting factor $W_{\rm r/p}= W_{\rm sign}|W_{\rm (r/p)hmc}|$ at the measurement stage to recover the target theory \cite{Luscher:2012av,Mohler:2020txx, Kuberski:2023zky}. Here, $W_{\rm sign}=\operatorname{sign}(\det D)$ is the discrete signature of the fermion determinant, whose reliable determination is essential whenever the fermion determinant is not manifestly positive. Recent strategies employ rational hybrid Monte Carlo (RHMC) approximation with a sufficiently large number of poles to render the associated reweighting factors $|W_{\rm rhmc}|$ effectively unity at the target precision \cite{Colquhoun:2024jzh}. In either case, whether the rational-approximation error is corrected through a compensating Metropolis accept-reject step \cite{JLQCD:2001ucs} or rendered trivial by employing a sufficiently high-degree RHMC approximation \cite{Colquhoun:2024jzh}, $W_{\rm sign}$ still requires explicit evaluation. 
	
	So far, two complementary approaches have been employed in the literature to determine the fermion determinant sign associated with Wilson fermions. The sign of $\det(D)$ is determined by identifying $N_{-}$ from the variation of low-lying eigenvalues of the Hermitian operator $Q=\gamma_5 D$ (note that $\det(D)=\det(Q)$) as a function of the fermion mass, commonly referred to as the spectral flow method \cite{Edwards:1998sh,Farchioni:2007dw,Mohler:2020txx, DellaMorte:2023ylq,Francis:2023gcm}. An alternative strategy is to work directly with the non-Hermitian Wilson-Dirac matrix operator and extract information on the number of negative real eigenvalues, if any exist \cite{Neff:2001xc,Bergner:2011zp,Ali:2018dnd}. The spectral information has also been exploited in determining the Pfaffian sign in lattice $\mathcal{N}=1$ supersymmetric Yang-Mills theory \cite{Ali:2018dnd,Steinhauser:2020zth}. Such direct approaches have traditionally relied on iterative eigensolvers, often supplemented by polynomial transformations \cite{Bergner:2011zp}, to resolve the low-lying real spectrum. 
	
	In this work, we exploit a contour-integral-based spectral projection to target directly the spectral region that determines the determinant signature, rather than resolving a broad low lying spectral region. The contour integral is supplemented by an adapted deflated accelerated solver for shifted linear systems and singular-value exclusions to reliably set spectral bounds and contour boundaries, which provides an efficient and systematic setup to isolate negative real eigenvalues of the non-Hermitian Wilson-Dirac operator. By restricting the eigensolver to the exact domain that governs the signature of the fermion determinant, our proposed method bypasses the computational overhead of resolving unwanted physical modes. Beyond the determination of the sign of the fermion determinant, the proposed procedure provides a systematic strategy with potential computational advantages for isolating physically relevant eigenmodes in large-scale non-Hermitian eigenvalue problems. Contour-integral eigensolvers have previously been applied to compute extremal eigenvalues in large-scale systems \cite{7765138}, low-lying eigenvalues of the non-Hermitian Wilson--Dirac operator in lattice QCD \cite{Suno:2016yrs,Futamura:2014kga}, and the sign function for the chiral overlap operator \cite{Nagai:2022fat}. In these studies, Sakurai-Sugiura methods were employed to extract eigenvalues within a prescribed region of the complex plane, whereas we utilize contour-integral-based spectral projection as a targeted filter for the specific subset of the Wilson–Dirac spectrum that determines the fermion-determinant sign. 
	
	The rest of the article is organized as follows. We first discuss the contour-integral spectral projector for general non-Hermitian matrices and its implementation, including the numerical procedures and criteria employed. We then present results from applying the proposed procedure to a diverse set of configurations from lattice QCD ensembles generated by the CLS consortium, followed by a summary.
	
	\section{Contour projection method}
	Any diagonalizable nonnormal matrix can be decomposed as $A(\in \mathbb{C}^{n \times n})=R\Lambda L^{\dagger}$, with $R=(r_1 ~r_2~r_3 ~...~ r_n)$ and $L= (l_1 ~l_2~l_3 ~...~ l_n)$ being the matrices made of right and left eigenvectors forming a biorthogonal basis satisfying $L^{\dagger}R=I$ and $\Lambda$ being the diagonal eigenvalue matrix. Let $Y \in \mathbb{C}^{n \times m}$ and $\tilde{Y} \in \mathbb{C}^{n \times m}$ be complex matrices, whose $m$ columns span the right and left trial subspace vectors, respectively. Employing an appropriately designed contour $\Gamma$, one can define the right projected subspace as
	\beq
	Q = \frac{1}{2\pi i} \oint_\Gamma (zI - A)^{-1} Y dz = P_\Gamma Y,
	\eeq{rpss}
	and the left projected subspace as
	\beq
	\tilde{Q} = \frac{-1}{2\pi i} \oint_{\Gamma} (\bar{z}I - A^\dagger)^{-1} \tilde{Y} d\bar{z} = P_\Gamma^\dagger \tilde{Y}.
	\eeq{lpss}
	See Appendix \ref{app:RSPF} for a detailed derivation. Utilizing the equations above, one can define the reduced matrix $A_Q \in \mathbb{C}^{m \times m}$ and the reduced overlap matrix $B_Q \in \mathbb{C}^{m \times m}$ as 
	\beq
	A_Q = \widetilde{Q}^\dagger A Q \mbox{\qquad and \qquad} B_Q = \widetilde{Q}^\dagger Q.
	\eeq{reduced}
	The target spectral information of the original matrix $A$ can then be approximated by the eigensolutions of the reduced generalized eigenvalue problem 
	\beq 
	A_Q w_i =  \lambda_iB_Q w_i,
	\eeq{redgevp}
	where $\lambda_i$ and $w_i$ are the eigenvalue and eigenvector of the reduced system, respectively. The corresponding full $n$-dimensional approximate right eigenvectors of $A$ can be constructed as $x_i = Qw_i$.
	
	To extract the negative real solutions of the Wilson-Dirac operator, we choose to work with an elliptical contour symmetric about real axis in the eigenvalue Argand plane defined as 
	\beq
	z(\theta) =E_c + a\cos{\theta} +i b\sin{\theta} \mbox{~ where ~} 0 < \theta <2\pi,
	\eeq{ellcon}
	and $E_c,~a,~b$ are the centre, semi-major axis, and semi-minor axis, respectively. By construction, the conjugation symmetry of the contour is consistent with the $\gamma_5$-Hermiticity of the Wilson-Dirac operator and the corresponding complex-conjugate symmetry of its spectrum. In terms of $\theta$, $P_{\Gamma}$ can be rewritten as  
	\beq
	P_\Gamma = \frac{1}{2\pi i}\int_{0}^{2\pi}(z(\theta)I- A)^{-1}z'(\theta)d\theta, 
	\eeq{ellproj}
	where $z'=dz/d\theta$. Numerically, we approximate the contour integral with a finite quadrature rule. To this end, we introduce $N_q$ quadrature nodes symmetrically placed around the real axis at $\theta_j= (j-\frac{1}{2}) 2\pi/N_q$ as defined in Ref. \cite{guettel2014zolotarevquadraturerulesload}. Then the approximated spectral projector can be expressed in terms of the quadrature nodes as 
	\beq
	\hat P_\Gamma\simeq\sum_{j=1}^{N_q}
	\omega_j(z_j I- A)^{-1} \mbox{~ where \qquad} \omega_j=\frac{z'(\theta_j)}{iN_q}
	\eeq{quadrature_projector}
	is the corresponding weight of the $j^{th}$ quadrature node. The resulting operator is only an approximation to the exact spectral projector, and in the limit $N_q\rightarrow\infty$, the rational projector gives the exact value of the continuum spectral projector, $\lim_{N_q\rightarrow\infty}\hat P_{\Gamma}=P_{\Gamma}$.
	
	Owing to the $\gamma_5$-Hermiticity of the Wilson-Dirac operator, its left and right eigensystems are related to each other. Following a little algebra, one can observe that if 
	\beq
	Ar_k = \lambda_k r_k \mbox{\qquad then \qquad} r^\dagger_k \gamma_5 A= \lambda^*_kr^\dagger_k \gamma_5
	\eeq{left_to_right}
	where $r^\dagger_k \gamma_5$ acts as left eigenvector associated with the conjugated eigenvalue $\lambda_k^*$. Thus, for the Wilson-Dirac operator, the relevant spectral information may be extracted from the right projected subspace and does not require an independent determination of the left eigensystem.  
	\section{Implementation and strategies \label{sec:ImpStr}}
	
	With even-odd decomposition of $D$, $\det(D)=\det(R_o)\det(\hat D)$, where $\hat D=R_e-D_{eo}R_o^{-1}D_{oe}$ is the even Schur complement \cite{DeGrand:1988vx,DeGrand:2006zz}, which inherits the $\gamma_5$-Hermiticity property from $D$. Provided $\det(R_o)>0$ (which is observed to be the case in all configurations we investigate, see also Ref. \cite{Mohler:2020txx}), the $W_{\operatorname{sign}}$ is exclusively determined by $\hat D$. Thus $W_{\operatorname{sign}}$ is evaluated by extracting negative-real eigenvalues of $\hat D$. To this end, we introduce a contour spectral projection $\hat P_{\Gamma}Y$ that proceeds through solving the shifted linear system 
	\beq
	(z_j I-\hat D)q_j=Y
	\eeq{shifted_system}
	at each quadrature node $z_j$. Using $q_j$ and $\omega_j$ (defined in Eq. \ref{eq:quadrature_projector}), the approximate projected subspace can be evaluated as $\hat Q=\sum_{j=1}^{N_q}\omega_j q_j$. We utilize the FEAST package~\cite{Kestyn2016} to realize the trial subspace vectors and the finite quadrature approximation. The trial-subspace dimension $m$ is chosen to exceed the expected number of eigenvalues enclosed by the contour. If the subspace becomes fully saturated during the FEAST iteration, the calculation must be repeated with a larger $m$. We observe that the first FEAST iteration of the approximated contour projection itself provides a useful indication of whether a negative-real mode is present inside the chosen contour. 
	
	To solve for $q_j$ at every quadrature node $z_j$, we have utilized an adapted $\mathtt{DFL\text{-}SAP}$-$\mathtt{GCR}$ solver for linear shifted systems in $\hat D$, constructed within the $\mathtt{openQCD}$ framework \cite{openQCD}. The existing solver in Ref. \cite{openQCD} cannot be applied directly to determine $q_j$ for two reasons. Firstly, the solver acts on the full Wilson-Dirac operator $D$ and not exclusively on $\hat D$. Secondly, and most importantly, the even Schur complement of the shifted full Wilson–Dirac operator ($\widehat{D-zI}$) is not equal to the shifted even Schur complement of the Wilson-Dirac operator $\hat D-zI$. The former can be implemented using the existing solver, whereas the latter, which is required in our case, necessitates an appropriate adaptation. Hence we have modified $\mathtt{DFL\text{-}SAP}$-$\mathtt{GCR}$ solver to act exclusively on the shifted even Schur complement $\hat D-zI$, which we refer to as shifted-$\mathtt{DFL\text{-}SAP}$-$\mathtt{GCR}$. This is observed to be particularly effective for the more demanding cases of ill-conditioned shifted systems. To ensure the solutions of $\hat D$ approximated from the solutions of the corresponding reduced system are sufficiently accurate, they are further passed through residual tolerance tests defined for the full lattice as well as tolerance tests for the imaginary part to filter out the real solutions.
	
	One critical subtlety is that the contour-integral-based spectral projector itself does not guarantee the absence of eigenvalues farther to the left of the left hand side extremum of the contour $X_L$. Consequently, if one or more negative real eigenvalues lie to the left of $X_L$, they will be excluded from the projected subspace and may therefore be missed. A first approximation for $X_L$ covering the lower bound in $\hat D$ spectrum can be made from the Rayleigh quotient $y^{\dagger}\hat H y=\mbox{Re}(y^{\dagger}\hat Dy)$ of the Hermitian part of $\hat D$, $\hat H = 0.5(\hat D + \hat D^\dagger)$. It naturally provides a lower bound on the real part of the solutions of $\hat D$, $\operatorname{min}_i \mbox{Re}(\lambda_i(\hat D)) \ge \vartheta_{min}(\hat H)$, where $y$ is the eigenvector corresponding to the lowest eigenvalue $\vartheta_{min}$ of $\hat H$. We estimate $\vartheta_{\min}(\hat H)$ using a coarse Hermitian Krylov solver, obtaining an approximate value $\tilde\vartheta_{\min}$, which is further rounded downwards to the second decimal to account for any remnant numerical uncertainties. For all the configurations considered here, we observe that $\tilde\vartheta_{\min}$ typically lies substantially farther to the left than the leftmost eigenvalues encountered in an extended contour search or estimated using other procedures. Hence, using it directly as $X_L$ would result in an unnecessarily large integration domain, requiring an increased number of quadrature nodes and potentially more stringent convergence tolerances for the shifted linear systems. To further refine $X_L$ to a more pragmatic estimate, we proceed with the observation that at any point $s_1$, the smallest singular value for the shifted system 
	\beq
	\sigma_{\min}\left(\hat D - s_1 I\right) = \sigma_{1,\min}\leq |\lambda-s_1|,
	\eeq{SVDconstraint}
	where $\lambda$ is an eigenvalue of $\hat D$ \cite{golub2013matrix}. Then the open interval $\varepsilon_{SV}=(s_1-\sigma_{1,\min},\;s_1+\sigma_{1,\min})$ would be free of eigenvalues of $\hat D$. This property can be exploited to identify an exclusion interval for the eigenvalues of $\hat D$ on a given configuration about $s_1$, if the exact value $\sigma_{1,\min}$ is known. 
	
	This observation motivates a more pragmatic choice for $X_L$ by combining singular-value estimates of the shifted systems $\hat D-s_i I$ evaluated at a set of suitably chosen points $s_i$. We introduce a sequence of ordered points along the real axis $\tilde\vartheta_{\min}=s_1<s_2<\cdots<\tilde X_L$ ($\tilde X_L$ is the initial user-provided value) to construct a contiguous exclusion interval extending from the left-most point $s_1$. The practical value of $X_L$ is then determined by the right-most endpoint of the connected component of the union of these exclusion intervals that contains $s_1$. Thus, if the exclusion intervals associated with all the chosen $s_i$ overlap to cover the interval $[s_1,\tilde X_L]$, the initially specified value $X_L=\tilde X_L$ is retained. If the exclusion ceases to be contiguous at $s_j$, the construction is terminated at the right boundary of the preceding exclusion interval $X_L=s_{j-1}+\sigma_{j-1,\min}$, with an additional rounding downward to the second decimal place to account for any numerical uncertainties. If the resulting exclusion regions contain gaps, additional shift points $s_i$ may be introduced within these gaps to refine the exclusion region. In this way, the singular-value information at a finite set of strategically chosen shifts provides a systematic and computationally inexpensive means of determining a conservative value of $X_L$, while avoiding an over-conservative choice based on $\tilde\vartheta_{\min}$. A few examples of this procedure, along with relevant details, are provided in Appendix \ref{app:SVEP}.
	
	\section{Results and discussion \label{sec:ResDis}}
	
\begin{figure}[htb]
		\includegraphics[width=\linewidth]{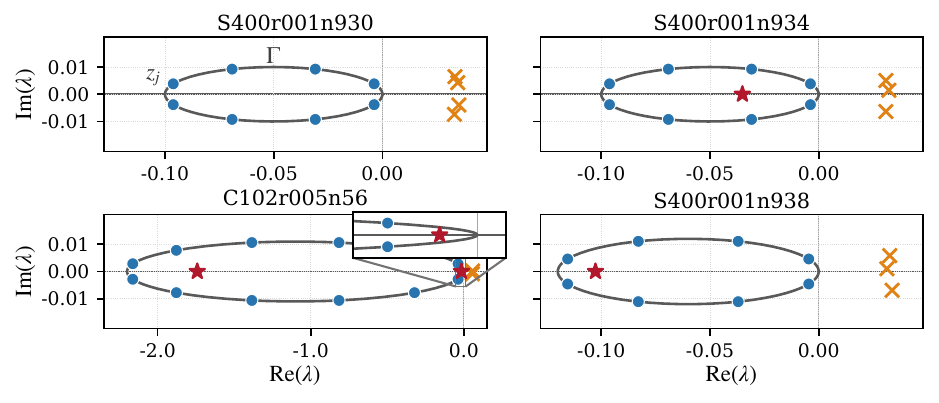}
		\caption{The Argand plane of the $\hat D$ eigenvalues for four representative configurations, one without any negative solutions \texttt{S400r001n930}, one with a shallow negative real solution  \texttt{S400r001n934}, one with a deep negative real solution \texttt{S400r001n938}, and one with two negative real solutions differing by two orders of magnitude \texttt{C102r005n56}.}
		\label{fig:S400_feast_spectrum}
	\end{figure}

	We have tested and verified that our proposed procedure reproduces all negative real solutions of $\hat D$ up to machine precision, and hence the fermion determinant signature, on a diverse subset of configurations from a number of lattice QCD ensembles generated by the CLS consortium \cite{Bruno:2014jqa}. In Appendix \ref{app:CLS}, we provide the details of the configurations studied here, and in this section, we present the results and observations from these tests. Throughout the tests, the elliptical contour $\Gamma$ is chosen to span the real-axis interval $[X_L,10^{-4}]$, where $X_L$ is chosen based on the singular value exclusion procedure discussed earlier, with the semi-minor to semi-major axis ratio fixed at $b/a=0.2$. In Fig. \ref{fig:S400_feast_spectrum}, we present the solutions obtained for a representative set of configurations with different properties. \texttt{S400r001n930} does not have any negative real solutions, whereas \texttt{S400r001n934}/\texttt{S400r001n938} has a shallow/deep negative real solution, and \texttt{C102r005n56} has two negative real solutions with magnitudes differing by two orders of magnitude. The solid black ellipse and the blue filled circles denote the chosen contour $\Gamma$ and the corresponding quadrature nodes $z_j$, respectively. Red stars denote $\hat D$ eigenvalues lying inside $\Gamma$, while the orange crosses outside denote representative solutions obtained from the FEAST iteration that lie close outside $\Gamma$. Note that $X_L=\tilde X_L$ applies to cases with no or shallow negative real solutions, whereas $|X_L|>|\tilde X_L|$ is chosen appropriately based on the singular value exclusion criteria for cases where there are deep negative real solutions. In Appendix \ref{app:SVEP}, we provide details of the singular value exclusions used to arrive at the final $X_L$ choices for each of these four configurations. Provided that a sufficient number of quadrature nodes are employed and that the modes are enclosed by $\Gamma$, the distance of the solutions from the origin is immaterial to their identification. The 
	computational overhead associated with choosing an appropriate elliptical contour and the number of quadrature nodes is less than a few percent of the entire evaluation. 
	
	A minimal choice of $N_q=6$, with the default nodal position assignment in FEAST \cite{FEASTUserGuide} (see Eq. \ref{eq:quadrature_projector}), and a shifted-solver tolerance of $10^{-3}$ was found to be sufficient for most of the configurations we have investigated. We choose a conservative value $N_q=8$ together with a shifted-solver tolerance of $10^{-3}$ throughout our study. We observe that all negative-real eigenvalues are already resolved after the second FEAST iteration for all configurations we have tested. Similarly, increasing the number of quadrature nodes from 8 to 12 does not lead to an appreciable change in the resolved eigenvalues, except in the special case of \texttt{C102r005n56}. Since the modes are successfully resolved after the second FEAST iteration, we employ eight quadrature nodes and two FEAST iterations for all configurations. We observe the same stability across the ensembles and configurations tested in this work, indicating that these parameters are sufficient for a reliable determination of all the negative real modes. 
	
	A nuance can arise when there are multiple eigenvalues that differ by more than an order of magnitude, thus requiring a contour with a large semi-major axis compared to the magnitude of the eigenvalue closer to the origin, as was the case for \texttt{C102r005n56}. In this case, we choose $N_q=12$, along with a tighter shifted-solver tolerance of $10^{-5}$ and $b/a \sim 0.01$, to faithfully capture both solutions. FEAST also provides a custom-contour facility, allowing users to specify the quadrature nodes $z_i$ and weights $\omega_i$ arbitrarily \cite{FEASTUserGuide}. One could also utilize this feature to span the quadrature nodes in a user-defined way, although we did not use this for our purposes. Despite the increased number of quadrature nodes and stringent tolerance, the observed computational overhead remained modest relative to the $N_q=8$ case.
	
	The dominant computational cost in the proposed framework arises from solving the shifted linear systems in Eq.~(\ref{eq:shifted_system}) at each quadrature node $z_j$. Consequently, for a fixed subspace dimension and solver tolerance, the total cost is controlled primarily by the dimension of $\hat D$ (the lattice half volume), the number of quadrature nodes, and the number of iterations required by the shifted linear solver. The latter also depend on the position of the quadrature nodes relative to the eigenvalue spectrum of $\hat D$. We provide the wall times for various configurations studied in Appendix \ref{app:CLS}.
	
	We close this section by emphasizing the scope of the proposed approach. It is particularly useful when the objective is to establish a systematically refined spectral boundary or to exclude regions of the complex plane without relying solely on the eigenvalues explicitly returned by an eigen-solver. For the non-Hermitian Wilson-Dirac operator, which can have both real and complex eigenvalues, the notion of a lowest eigenvalue is subject to the spectral ordering adopted. Conventional eigensolvers are designed to extract the set of extremal eigenvalues according to a chosen criterion, whereas the proposed procedure, exploiting contour integrals and singular value-based spectral bounds, is designed to perform spectral filtering within a specified region. Practically, the shifted systems associated with different quadrature nodes can be treated independently and are well suited to straightforward parallelization. Furthermore, using the shifted-$\mathtt{DFL\text{-}SAP\text{-}GCR}$ solver improves the computational efficiency of the linear algebra involved. In this sense, the present procedure brings together several complementary algorithmic components to provide a controlled and region-selective framework for determining spectral information.
	
	\section{Summary \label{sec:Sum}}
	
	In this work, we propose a numerical procedure to determine the signature of a rationally approximated Wilson-Dirac fermion determinant based on a contour-integral-based spectral projector by identifying all its negative-real solutions. Complementing the contour-integral-quadrature-based spectral projection with the shifted-$\mathtt{DFL\text{-}SAP\text{-}GCR}$ solver and singular-value exclusion criteria for spectral bound assignments allows all the negative-real modes to be distinguished reliably and provides a direct determination of the determinant sign. We have tested and verified the proposed procedure on a diverse set of CLS gauge configurations in spectral regions where negative as well as positive determinants occur. Our results establish contour projection as a practical and complementary alternative to conventional spectral-flow methods for determining the Wilson-fermion determinant sign. The method is particularly attractive when following individual modes through a mass scan becomes inconvenient or when direct access to the spectrum at the target quark mass is desirable. An important direction for future work is to extend and systematically benchmark the method over a wider range of lattice spacings, volumes, quark masses, and gauge-field ensembles, including cases with increasingly dense or closely spaced real modes. 
	
	A potential application of our procedure beyond the present work is in RHMC simulations, where one could devise a systematic means of identifying, validating, and potentially optimizing the assumed spectral ranges, which control the accuracy of the rational approximation. In particular, the singular-value machinery in our framework could be adapted to certify the lower spectral bound relevant to the RHMC approximation, thereby identifying exceptional configurations whose low modes fall outside the assumed spectral interval, thus suggesting a need for dedicated reweighting or low-mode treatment. From a broader perspective, the proposed procedure provides a systematic strategy for isolating and validating physically relevant eigenmodes in prescribed regions of a dense complex spectrum, which may be useful in stability analyses, non-Hermitian quantum systems, and large-scale eigenvalue problems arising in computational physics.
	
	\section{Acknowledgments}
	
	We thank the members of the Coordinated Lattice Simulations (CLS) consortium, Gunnar Bali, Jeremy Green, Simon Kuberski, and Wolfgang S\"oldner, for sharing their gauge ensembles \cite{Bruno:2014jqa}. In particular, we thank Simon Kuberski and Wolfgang S\"oldner for valuable discussions and for sharing the strange reweighting factors. We are grateful to our collaborators Nilmani Mathur, Ram Prakash, and Tanishk Shrimal and other participants of the Monsoon Hadrons Workshop (\href{https://scitalks.tifr.res.in/event/9330/}{https://scitalks.tifr.res.in/event/9330/}) for various discussions. We have utilized a modified version of the $\mathtt{DFL\text{-}SAP\text{-}GCR}$ solver for the inversion of the Wilson-Dirac operator, as well as other utilities from \texttt{openQCD} \cite{openQCD}. We also acknowledge the use of the FEAST package \cite{Kestyn2016,FEASTUserGuide} for constructing the trial subspace vectors and contour integral evaluation using a finite quadrature rule. We acknowledge support from the Institute of Mathematical Sciences and the use of the HPC facility at IMSc Chennai. M.P. gratefully acknowledges support from the Department of Science and Technology, India, ANRF(SERB earlier) Start-up Research Grant No. SRG/2023/001235 and the Department of Atomic Energy, India.
	
	\bibliographystyle{elsarticle-num}
	\bibliography{references}

@article{Kuberski:2023zky,
    author = "Kuberski, Simon",
    title = "{Low-mode deflation for twisted-mass and RHMC reweighting in lattice QCD}",
    eprint = "2306.02385",
    archivePrefix = "arXiv",
    primaryClass = "hep-lat",
    reportNumber = "MITP-23-021",
    doi = "10.1016/j.cpc.2024.109173",
    journal = "Comput. Phys. Commun.",
    volume = "300",
    pages = "109173",
    year = "2024"
}

@misc{openQCD,
  author       = {L{\"u}scher, Martin and Schaefer, Stefan},
  title        = {{openQCD}: Simulation Programs for Lattice QCD},
  howpublished = {\url{https://luscher.web.cern.ch/luscher/openQCD/}},
  note         = {Version 2.4.2},
  year         = {2024}
}

@article{Luscher:2012av,
    author = "Luscher, Martin and Schaefer, Stefan",
    title = "{Lattice QCD with open boundary conditions and twisted-mass reweighting}",
    eprint = "1206.2809",
    archivePrefix = "arXiv",
    primaryClass = "hep-lat",
    reportNumber = "CERN-PH-TH-2012-161",
    doi = "10.1016/j.cpc.2012.10.003",
    journal = "Comput. Phys. Commun.",
    volume = "184",
    pages = "519--528",
    year = "2013"
}

@article{Bruno:2014jqa,
    author = "Bruno, Mattia and others",
    title = "{Simulation of QCD with N$_{f} =$ 2 $+$ 1 flavors of non-perturbatively improved Wilson fermions}",
    eprint = "1411.3982",
    archivePrefix = "arXiv",
    primaryClass = "hep-lat",
    reportNumber = "DESY-14-216, FTUAM-14-48, HIM-2014-01, HU-EP-14-51, MITP-14-091, SFB-CPP-14-89, IFT-UAM-CSIC-14-117",
    doi = "10.1007/JHEP02(2015)043",
    journal = "JHEP",
    volume = "02",
    pages = "043",
    year = "2015"
}

@article{Kestyn2016,
  author  = {Kestyn, James and Polizzi, Eric and Peter Tang, Ping Tak},
  title   = {{FEAST} Eigensolver for Non-Hermitian Problems},
  journal = {SIAM Journal on Scientific Computing},
  volume  = {38},
  number  = {5},
  pages   = {S772--S799},
  year    = {2016},
  doi     = {10.1137/15M1026572}
}

@misc{FEASTUserGuide,
    title = {FEAST Eigenvalue Solver v4.0 User Guide},
    author = {Polizzi, Eric and Kestyn, James and Gavin, Brendan and Güttel, Stefan and Brenneck, Julien},
    howpublished = {\url{http://www.feast-solver.org}},
    year = {2020},
}

@article{Mohler:2020txx,
    author = "Mohler, Daniel and Schaefer, Stefan",
    title = "{Remarks on strange-quark simulations with Wilson fermions}",
    eprint = "2003.13359",
    archivePrefix = "arXiv",
    primaryClass = "hep-lat",
    reportNumber = "DESY-20-041, MITP/20-010, DESY 20-041 ; MITP/20-010",
    doi = "10.1103/PhysRevD.102.074506",
    journal = "Phys. Rev. D",
    volume = "102",
    number = "7",
    pages = "074506",
    year = "2020"
}

@article{RCstar:2022yjz,
    author = {Bushnaq, Lucius and Campos, Isabel and Catillo, Marco and Cotellucci, Alessandro and Dale, Madeleine and Fritzsch, Patrick and L{\"u}cke, Jens and Krsti{\'c} Marinkovi{\'c}, Marina and Patella, Agostino and Tantalo, Nazario},
    collaboration = "RCstar",
    title = "{First results on QCD+QED with C$^{*}$ boundary conditions}",
    eprint = "2209.13183",
    archivePrefix = "arXiv",
    primaryClass = "hep-lat",
    reportNumber = "HU-EP-22/29-RTG",
    doi = "10.1007/JHEP03(2023)012",
    journal = "JHEP",
    volume = "03",
    pages = "012",
    year = "2023"
}

@article{DeGrand:1988vx,
    author = "DeGrand, Thomas A.",
    title = "{A Conditioning Technique for Matrix Inversion for Wilson Fermions}",
    reportNumber = "COLO-HEP-176",
    doi = "10.1016/0010-4655(88)90180-4",
    journal = "Comput. Phys. Commun.",
    volume = "52",
    pages = "161--164",
    year = "1988"
}

@article{Sint:1993un,
    author = "Sint, Stefan",
    title = "{On the Schrodinger functional in QCD}",
    eprint = "hep-lat/9312079",
    archivePrefix = "arXiv",
    reportNumber = "DESY-93-165, DESY-93--165",
    doi = "10.1016/0550-3213(94)90228-3",
    journal = "Nucl. Phys. B",
    volume = "421",
    pages = "135--158",
    year = "1994"
}

@book{golub2013matrix,
  title={Matrix Computations},
  author={Golub, Gene H. and Van Loan, Charles F.},
  edition={4th},
  year={2013},
  publisher={Johns Hopkins University Press},
  address={Baltimore},
  isbn={978-1421407944}
}

@article{Colquhoun:2024jzh,
    author = "Colquhoun, B. and Francis, A. and Hudspith, R. J. and Lewis, R. and Maltman, K. and Parrott, W. G.",
    title = "{Improved analysis of strong-interaction-stable doubly bottom tetraquarks on the lattice}",
    eprint = "2407.08816",
    archivePrefix = "arXiv",
    primaryClass = "hep-lat",
    doi = "10.1103/PhysRevD.110.094503",
    journal = "Phys. Rev. D",
    volume = "110",
    number = "9",
    pages = "094503",
    year = "2024"
}

@article{JLQCD:2001ucs,
    author = "Aoki, S. and others",
    collaboration = "JLQCD",
    title = "{Polynomial hybrid Monte Carlo algorithm for lattice QCD with odd number of flavors}",
    eprint = "hep-lat/0112051",
    archivePrefix = "arXiv",
    reportNumber = "KEK-CP-120, UTCPP-P-121, UTHEP-455",
    doi = "10.1103/PhysRevD.65.094507",
    journal = "Phys. Rev. D",
    volume = "65",
    pages = "094507",
    year = "2002"
}

@article{Bergner:2011zp,
    author = "Bergner, G. and Wuilloud, J.",
    title = "{Acceleration of the Arnoldi method and real eigenvalues of the non-Hermitian Wilson-Dirac operator}",
    eprint = "1104.1363",
    archivePrefix = "arXiv",
    primaryClass = "hep-lat",
    reportNumber = "MS-TP-11-07",
    doi = "10.1016/j.cpc.2011.10.007",
    journal = "Comput. Phys. Commun.",
    volume = "183",
    pages = "299--304",
    year = "2012"
}

@article{Neff:2001xc,
    author = "Neff, H.",
    editor = "Muller-Preussker, M. and Bietenholz, Wolfgang and Jansen, K. and Jegerlehner, F. and Montvay, I. and Schierholz, G. and Sommer, R. and Wolff, U.",
    title = "{Efficient computation of low lying eigenmodes of nonHermitian Wilson-Dirac type matrices}",
    eprint = "hep-lat/0110076",
    archivePrefix = "arXiv",
    doi = "10.1016/S0920-5632(01)01926-0",
    journal = "Nucl. Phys. B Proc. Suppl.",
    volume = "106",
    pages = "1055--1057",
    year = "2002"
}

@article{DellaMorte:2023ylq,
    author = {Della Morte, Michele and J{\"a}ger, Benjamin and Sannino, Francesco and Tsang, Justus Tobias and Ziegler, Felix P. G.},
    title = "{Spectrum of QCD with one flavor: A window for supersymmetric dynamics}",
    eprint = "2302.10514",
    archivePrefix = "arXiv",
    primaryClass = "hep-lat",
    reportNumber = "CERN-TH-2023-028",
    doi = "10.1103/PhysRevD.107.114506",
    journal = "Phys. Rev. D",
    volume = "107",
    number = "11",
    pages = "114506",
    year = "2023"
}

@article{Francis:2018xjd,
    author = "Francis, Anthony and Hudspith, Renwick J. and Lewis, Randy and Tulin, Sean",
    title = "{Dark Matter from Strong Dynamics: The Minimal Theory of Dark Baryons}",
    eprint = "1809.09117",
    archivePrefix = "arXiv",
    primaryClass = "hep-ph",
    reportNumber = "CERN-TH-2018-207",
    doi = "10.1007/JHEP12(2018)118",
    journal = "JHEP",
    volume = "12",
    pages = "118",
    year = "2018"
}

@article{Clark:2006fx,
    author = "Clark, M. A. and Kennedy, A. D.",
    title = "{Accelerating dynamical fermion computations using the rational hybrid Monte Carlo (RHMC) algorithm with multiple pseudofermion fields}",
    eprint = "hep-lat/0608015",
    archivePrefix = "arXiv",
    doi = "10.1103/PhysRevLett.98.051601",
    journal = "Phys. Rev. Lett.",
    volume = "98",
    pages = "051601",
    year = "2007"
}

@article{Frezzotti:1998eu,
    author = "Frezzotti, Roberto and Jansen, Karl",
    title = "{The PHMC algorithm for simulations of dynamical fermions: 1. Description and properties}",
    eprint = "hep-lat/9808011",
    archivePrefix = "arXiv",
    reportNumber = "CERN-TH-98-237, MPI-PHT-98-51",
    doi = "10.1016/S0550-3213(99)00321-1",
    journal = "Nucl. Phys. B",
    volume = "555",
    pages = "395--431",
    year = "1999"
}

@article{Edwards:1998sh,
    author = "Edwards, Robert G. and Heller, Urs M. and Narayanan, Rajamani",
    title = "{Spectral flow, chiral condensate and topology in lattice QCD}",
    eprint = "hep-lat/9802016",
    archivePrefix = "arXiv",
    reportNumber = "FSU-SCRI-98-15",
    doi = "10.1016/S0550-3213(98)00588-4",
    journal = "Nucl. Phys. B",
    volume = "535",
    pages = "403--422",
    year = "1998"
}

@article{Farchioni:2007dw,
    author = "Farchioni, F. and Montvay, I. and Munster, G. and Scholz, E. E. and Sudmann, T. and Wuilloud, J.",
    title = "{Hadron masses in QCD with one quark flavour}",
    eprint = "0706.1131",
    archivePrefix = "arXiv",
    primaryClass = "hep-lat",
    reportNumber = "DESY-07-078, MS-TP-07-14, BNL-HET-07-9",
    doi = "10.1140/epjc/s10052-007-0394-4",
    journal = "Eur. Phys. J. C",
    volume = "52",
    pages = "305--314",
    year = "2007"
}

@article{Ali:2018dnd,
    author = {Ali, Sajid and Bergner, Georg and Gerber, Henning and Giudice, Pietro and Montvay, Istvan and M{\"u}nster, Gernot and Piemonte, Stefano and Scior, Philipp},
    title = "{The light bound states of $\mathcal{N}=1$ supersymmetric SU(3) Yang-Mills theory on the lattice}",
    eprint = "1801.08062",
    archivePrefix = "arXiv",
    primaryClass = "hep-lat",
    reportNumber = "MS-TP-18-04, DESY 18-008, DESY-18-008",
    doi = "10.1007/JHEP03(2018)113",
    journal = "JHEP",
    volume = "03",
    pages = "113",
    year = "2018"
}

@article{Luscher:2011kk,
    author = "L{\"u}scher, Martin and Schaefer, Stefan",
    title = "{Lattice QCD without topology barriers}",
    eprint = "1105.4749",
    archivePrefix = "arXiv",
    primaryClass = "hep-lat",
    doi = "10.1007/JHEP07(2011)036",
    journal = "JHEP",
    volume = "07",
    pages = "036",
    year = "2011"
}

@misc{guettel2014zolotarevquadraturerulesload,
      title={Zolotarev Quadrature Rules and Load Balancing for the FEAST Eigensolver}, 
      author={Stefan Guettel and Eric Polizzi and Ping Tak Peter Tang and Gautier Viaud},
      year={2014},
      eprint={1407.8078},
      archivePrefix={arXiv},
      primaryClass={math.NA},
}

@book{DeGrand:2006zz,
    author = "DeGrand, Thomas and Detar, Carleton E.",
    title = "{Lattice methods for quantum chromodynamics}",
    year = "2006"
}

@article{Suno:2016yrs,
    author = "Suno, Hiroya and Nakamura, Yoshifumi and Ishikawa, K. I. and Kuramashi, Y. and Futamura, Yasunori and Imakura, Akira and Sakurai, Tetsuya",
    title = "{Eigenspectrum calculation of the non-Hermitian $O(a)$-improved Wilson-Dirac operator using the Sakurai-Sugiura method}",
    doi = "10.22323/1.251.0026",
    journal = "PoS",
    volume = "LATTICE2015",
    pages = "026",
    year = "2016"
}

@article{Nagai:2022fat,
    author = "Nagai, Yuki and Tomiya, Akio",
    title = "{Extensively parallelizable chiral fermion}",
    eprint = "2204.01583",
    archivePrefix = "arXiv",
    primaryClass = "hep-lat",
    month = "4",
    year = "2022"
}

@article{Futamura:2014kga,
    author = "Futamura, Yasunori and Hashimoto, Shoji and Imakura, Akira and Nagata, Keitaro and Sakurai, Tetsuya",
    title = "{A filtering technique for the temporally reduced matrix of the Wilson fermion determinant}",
    eprint = "1411.4262",
    archivePrefix = "arXiv",
    primaryClass = "hep-lat",
    reportNumber = "KEK-CP-314.-MSN-024",
    doi = "10.22323/1.214.0049",
    journal = "PoS",
    volume = "LATTICE2014",
    pages = "049",
    year = "2014"
}

@article{7765138,
  author={Li, Yongjie and Geng, Guangchao and Jiang, Quanyuan},
  journal={IEEE Transactions on Smart Grid}, 
  title={A Parallelized Contour Integral Rayleigh–Ritz Method for Computing Critical Eigenvalues of Large-Scale Power Systems}, 
  year={2018},
  volume={9},
  number={4},
  pages={3573-3581},
  doi={10.1109/TSG.2016.2635159}}

@article{Neff:2001zr,
    author = "Neff, H. and Eicker, N. and Lippert, T. and Negele, John W. and Schilling, K.",
    title = "{On the low fermionic eigenmode dominance in QCD on the lattice}",
    eprint = "hep-lat/0106016",
    archivePrefix = "arXiv",
    doi = "10.1103/PhysRevD.64.114509",
    journal = "Phys. Rev. D",
    volume = "64",
    pages = "114509",
    year = "2001"
}

@article{Steinhauser:2020zth,
    author = {Steinhauser, Marc and Sternbeck, Andr{\'e} and Wellegehausen, Bj{\"o}rn and Wipf, Andreas},
    title = "{$\mathcal{N}=1$ Super-Yang-Mills theory on the lattice with twisted mass fermions}",
    eprint = "2010.00946",
    archivePrefix = "arXiv",
    primaryClass = "hep-lat",
    doi = "10.1007/JHEP01(2021)154",
    journal = "JHEP",
    volume = "01",
    pages = "154",
    year = "2021"
}

@article{Francis:2023gcm,
    author = "Francis, Anthony and Cuteri, Francesca and Fritzsch, Patrick and Pederiva, Giovanni and Rago, Antonio and Shindler, Andrea and Walker-Loud, Andre and Zafeiropoulos, Savvas",
    title = "{Progress in generating gauge ensembles with Stabilized Wilson Fermions}",
    eprint = "2312.11298",
    archivePrefix = "arXiv",
    primaryClass = "hep-lat",
    doi = "10.22323/1.453.0048",
    journal = "PoS",
    volume = "LATTICE2023",
    pages = "048",
    year = "2024"
}
	
	\newpage 
	
	\appendix
	\section{Reisz spectral projection formulae \label{app:RSPF}}
	In this Appendix, we derive Eqs. \ref{eq:rpss} and \ref{eq:lpss} presented in the main draft. Unlike Hermitian matrices, diagonalizable nonnormal matrices possess distinct left and right eigenvectors that form a biorthogonal basis. Let $A \in \mathbb{C}^{n \times n}$ be a nonnormal diagonalizable matrix (for $\hat D$, $n=N_e=6L_0 \times L_1 \times L_2 \times L_3$). Then the matrices of left eigenvectors $L= (l_1 ~l_2~l_3 ~...~ l_n) $ and right eigenvectors $R= (r_1 ~r_2~r_3 ~...~ r_n)$ satisfy the eigenvalue equations
	\beq
	AR=\Lambda R \mbox{\qquad and \qquad} L^\dagger A = L^\dagger\Lambda,
	\eeq{biortho}
	where $\Lambda$ is the diagonal matrix of eigenvalues of $A$. The eigenvectors $l_i$ and $r_i$ are normalized to satisfy the biorthogonality condition $L^\dagger R =I$, or, component-wise, $l^\dagger_ir_j =\delta_{ij}$. Given this, for any complex scalar $z$ that is not an eigenvalue of $A$, one can write 
	\beq
	(zI-A)^{-1} = R(zI-\Lambda)^{-1}L^\dagger = \sum_{j=1}^n \frac{r_j l^\dagger_j}{z-\lambda_j}.
	\eeq{resolvent}
	If $z$ coincides with an eigenvalue of $A$, the resolvent $(zI-A)^{-1}$ is singular; if it lies sufficiently close to an eigenvalue, the corresponding shifted system becomes ill-conditioned. Utilizing this mode expansion, one can define the Riesz spectral projector $P_{\Gamma}$ using Cauchy's residue theorem as 
	\begin{equation}
		P_\Gamma = \frac{1}{2\pi i}\oint_\Gamma(zI- A)^{-1}dz = \sum_{j=1}^n r_j l^\dagger_j \left[\frac{1}{2\pi i}\oint_\Gamma \frac{dz}{z-\lambda_j}\right].
		\label{eq:contour_projector}
	\end{equation}
	If the contour $\Gamma$ encloses exactly $N_{in}$ eigenvalues, the integral filters out the exterior modes, and the projector reduces to
	\begin{equation}
		P_{\Gamma} =\sum_{j=1}^{N_{in}} r_jl^\dagger_j.
	\end{equation}
	
	To target a specific spectral region enclosed by the contour $\Gamma$, the Riesz spectral projector is applied to a trial subspace. Let $Y \in \mathbb{C}^{n \times m}$ and $\tilde{Y} \in \mathbb{C}^{n \times m}$ be complex matrices whose $m$ columns span the right and left trial subspaces, respectively. Then the right projected subspace is given by
	\beq
	Q = P_\Gamma Y = \frac{1}{2\pi i} \oint_\Gamma (zI - A)^{-1} Y dz = \sum_{j=1}^{N_{\text{in}}} r_j(l^\dagger_j Y) = \sum_{j=1}^{N_{\text{in}}} r_jc_j^T.
	\eeq{arpss}
	Similarly, the left projected subspace is given by
	\beq
	\tilde{Q} = P_\Gamma^\dagger \tilde{Y} = \frac{-1}{2\pi i} \oint_{\Gamma} (\bar{z}I - A^\dagger)^{-1} \tilde{Y} d\bar{z} = \sum_{j=1}^{N_{\text{in}}} l_j(r^\dagger_j \tilde{Y}) = \sum_{j=1}^{N_{\text{in}}} l_j\tilde{c}_j^T
	\eeq{alpss}
	Here, the factors $c_j^T$ and $\tilde{c}_j^T$ represent the overlaps of the enclosed eigenmodes of $A$ with the initial trial subspaces $Y$ and $\tilde Y$. 
	
	\clearpage 
	
	\section{Singular value exclusions and $X_L$ choices \label{app:SVEP}}
\begin{figure}[htbp]
		\centering
		\includegraphics[width=0.8\linewidth]{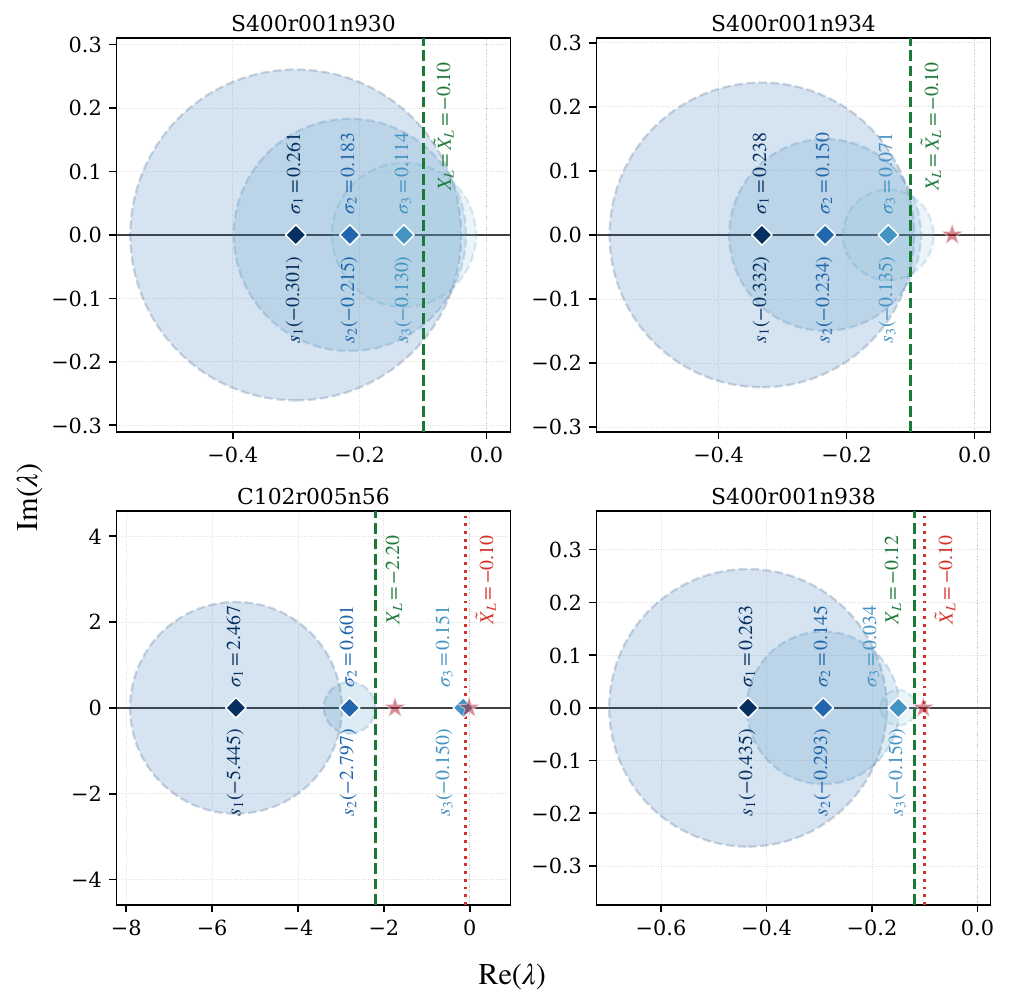}
		\caption{The Argand plane of the $\hat D$ eigenvalues for four representative configurations, demonstrating the singular value exclusion regions and the choice of $X_L$ in each of them. The location of negative real solutions are indicated by red stars.}
		\label{fig:SVEP}
	\end{figure}

	In this section, we provide examples of the search for a contiguous exclusion interval based on the singular values of the shifted system $\hat D-s_iI$ extending from the left-most point $s_1$ to the user-input $\tilde X_L$. We consider three points along the real axis $\tilde\vartheta_{\min}=s_1<s_2<s_3$ in the interval $[\tilde\vartheta_{\min},\tilde X_L)$ across all the configurations and identify the singular value exclusion regions to arrive at a pragmatic choice for $X_L$. The $s_i$ are chosen using the formula
	\begin{equation}
		s_i = \tilde\vartheta_{\min} +\frac{i-1}{n_p -1}\left[(\tilde X_L-\delta_{\mathrm{buff}}) -\tilde\vartheta_{\min}\right]
	\end{equation}
	where $n_p$ is the number of probe points and $\delta_{\mathrm{buff}}= \mathrm{min}\left[0.05,\mathrm{max(0.02,0.15|\tilde X_L-\tilde\vartheta_{\min}}|)\right]$. Here, 0.02 and 0.05 are the minimum and maximum absolute buffers, respectively and 0.15 is choosen as the relative buffer fraction.
	In Figure \ref{fig:SVEP}, we present this construction for the same configurations discussed in the main manuscript. For configurations \texttt{S400r001n930} and \texttt{S400r001n934}, $X_L$ was chosen to be the user-input value $\tilde X_L=-0.10$, as there are no negative real solutions in the former and a single shallow negative real solution in the latter. In the configuration \texttt{S400r001n938}, the presence of a relatively deep negative real solution pushed the singular value-based exclusion regions to set $X_L=-0.12$. Configuration \texttt{C102r005n56} possesses two negative real solutions, one that is shallow and one that is as large as $-1.74$, forcing the singular value exclusion procedure to set a conservative yet pragmatic value of $X_L=-2.20$.
	
	\newpage
	
	\section{CLS lattice QCD ensembles and configurations \label{app:CLS}}
	
	In Table \ref{tab:cnfg_CLS_scale}, we present the list of CLS lattice QCD configurations utilized in this work to test and verify the proposed procedure. The trial subspace dimension is chosen to be $m=4$, which we have observed to be sufficiently large for all tests performed in the present work. The ensemble list include those with open as well as periodic boundary conditions along the temporal extent, which are identified in the third column of Table \ref{tab:cnfg_CLS_scale}. For ensembles with open boundary conditions, the boundary constraints \cite{Luscher:2011kk, Sint:1993un} are enforced on the trial subspace $Y$. All the tests were carried out on Intel Xeon Platinum 8358 processors on our local computing cluster Kamet at IMSc Chennai. We present the computational resources utilized in the fifth column and the wall-times in the seventh and ninth columns for configurations with negative and positive fermion determinants, respectively. 
	\begin{table}[htbp]
		\centering
		\caption{The subset of lattice QCD configurations, from ensembles generated by the CLS consortium, used to test and benchmark our sign determination procedure using the contour-integral-based spectral projector. For each ensemble, at least one positive-sign and one negative-sign configuration are tested. The wall time per configuration (in seconds) is quoted in the seventh and ninth columns next to the configuration numbers. The ensemble \texttt{C102r005} is highlighted with an asterisk to indicate that the configurations studied are observed to have two negative-real solutions and thus have effectively a positive strange-quark determinant.}
		\label{tab:cnfg_CLS_scale}
		\setlength{\tabcolsep}{4pt}
		\begin{tabular}{lcccccccc}
			\hline\hline
			Ensemble ID & Geometry & Boundary & $m_K$ [MeV] & cores &
			\multicolumn{2}{c}{Negative Sign} &
			\multicolumn{2}{c}{Positive Sign} \\
			\cline{6-7}\cline{8-9}
			& & & & &
			Config. & Wall time [s] &
			Config. & Wall time [s] \\
			\hline\hline
			\texttt{U103r002}
			& $128\times24^3$ & obc & 420 & 192
			& 884 & 68 & 882 & 45 \\
			& & & & & 885 & 69 & 883 & 44 \\
			\cmidrule(lr){1-9}
			\texttt{U102r002}
			& $128\times24^3$ & obc & 440 & 192
			& 3140 & 64 & 3128 & 45 \\
			& & & & & 3146 & 64 & 3134 & 47 \\
			\cmidrule(lr){1-9}
			\texttt{H105r001}
			& $96\times32^3$ & obc & 460 & 192
			& 100 & 119 & 95 & 75 \\
			& & & & & 105 & 113 & 110 & 73 \\
			\cmidrule(lr){1-9}
			\texttt{S400r001}
			& $128\times32^3$ & obc & 440 & 128
			& 934 & 268 & 930 & 167 \\
			& & & & & 938 & 226 & 952 & 167 \\
			\cmidrule(lr){1-9}
			\texttt{C101r014}
			& $96\times48^3$ & obc & 470 & 192
			& 764 & 378 & 1838 & 260 \\
			& & & & & 1830 & 373 & & \\
			\cmidrule(lr){1-9}
			\texttt{N101r001}
			& $128\times48^3$ & obc & 460 & 192
			& 203 & 406 & 194 & 367 \\
			& & & & & 210 & 487 & 228 & 360 \\
			\cmidrule(lr){1-9}
			\texttt{N401r000}
			& $128\times48^3$ & obc & 460 & 192
			& 508 & 537 & 500 & 363 \\
			& & & & & 520 & 501 & 590 & 435 \\
			\cmidrule(lr){1-9}
			\texttt{A652r001}
			& $48\times24^3$ & pbc & 420 & 192
			& 1025 & 23 & 1015 & 20 \\
			& & & & & 1030 & 34 & & \\
			\cmidrule(lr){1-9}
			\texttt{X650r001}
			& $48\times48^3$ & pbc & 460 & 192
			& 203 & 207 & 195 & 146 \\
			& & & & & 210 & 193 & 200 & 141 \\
			\cmidrule(lr){1-9}
			\texttt{X150r001}
			& $128\times40^3$ & pbc & 460 & 128
			& 235 & 540 & 230 & 388 \\
			& & & & & 245 & 622 & 240 & 386 \\
			\hline
			\texttt{C102r005}$^{*}$
			& $96\times48^3$ & obc & 460 & 192
			& - & - & 56 & 498 \\
			& & & & & & & 72 & 323 \\
			& & & & & & & 73 & 392 \\
			\hline\hline
		\end{tabular}
	\end{table}
	
\end{document}